\documentclass[prl,twocolumn,showpacs,showkeys,groupaddress,preprintnumbers,floatfix]{revtex4-1}

\usepackage{amsmath}
\usepackage{dynlearn}
\usepackage{physics}
\usepackage{mathrsfs}
\usepackage{mathtools}

\begin{document}

\def\ourTitle{%
Islands in the Gap:\\
Intertwined Transport and Localization in\\
Structurally Complex Materials
}

\def\ourAbstract{%
Localized waves in disordered one-dimensional materials have been studied for
decades, including white-noise and correlated disorder, as well as
quasi-periodic disorder. How these wave phenomena relate to those in
crystalline (periodic ordered) materials---arguably the better understood
setting---has been a mystery ever since Anderson discovered disorder-induced
localization.  Nonetheless, together these revolutionized materials science and
technology and led to new physics far beyond the solid state. We introduce a
broad family of structurally complex materials---chaotic crystals---that
interpolate between these organizational extremes---systematically spanning
periodic structures and random disorder. Within the family one can tune the
degree of disorder to sweep through an intermediate structurally disordered
region between two periodic lattices. This reveals new transport and
localization phenomena reflected in a rich array of energy-dependent
localization degree and density of states. In particular, strong localization
is observed even with a very low degree of disorder. Moreover, markedly
enhanced localization and delocalization coexist in a very narrow range of
energies. Most notably, beyond the simply smoothed bands found in previous
disorder studies, islands of transport emerge in band gaps and sharp band
boundaries persist in the presence of substantial disorder. Finally, the family
of materials comes with rather direct specifications of how to assemble the
requisite material organizations.
}

\def\ourKeywords{%
Anderson localization, hidden Markov process, structured disorder,
$\epsilon$-machines, computational mechanics
}

\hypersetup{
  pdfauthor={James P. Crutchfield},
  pdftitle={\ourTitle},
  pdfsubject={\ourAbstract},
  pdfkeywords={\ourKeywords},
  pdfproducer={},
  pdfcreator={}
}

\author{Xincheng~Lei}
\email{xclei@ucdavis.edu}

\author{Dowman~P.~Varn}
\email{dpv@complexmatter.org}

\author{James~P.~Crutchfield}
\email{chaos@ucdavis.edu}

\affiliation{Complexity Sciences Center and Physics Department,
University of California at Davis, One Shields Avenue, Davis, CA 95616}

\date{\today}
\bibliographystyle{unsrt}

%%%%%%%%%%%%%%%%%%%%%%%%%%%%%%%%%%%%%%%%%%%%%%%%%%%%%%%%%%%%%%%%%%%%%%%%%%%%%%%

\title{\ourTitle}

\begin{abstract}

\ourAbstract

\end{abstract}



\preprint{\sfiwp{17-07-XXX}}
\preprint{\arxiv{1707.XXXXX [cond-mat.stat-mech]}}

\title{\ourTitle}
\date{\today}
\maketitle

\setstretch{1.1}

The quantum mechanics of wave phenomena in crystals led to unprecedented
successes in understanding materials, from conductors to semiconductors and
insulators. Technologically harnessing the scientific advances, in turn, led to
a scale of social impacts that cannot be understated. However, the vast
majority of materials are not so ordered. Understanding the properties of
disordered materials was greatly advanced when it was discovered how structural
disorder breaks crystalline wave-transport phenomena.

To date, analyzing the material properties arising at these two structural
extremes requires distinct concepts and methods. Indeed, regular lattices
lead to wavefunction states that extend over the whole system, whereas 
random lattices produce localized states at all energies. Is there an
overarching theory? And, what about the existence of materials that are
intermediate, neither perfectly ordered nor utterly random? What is the
character of their quantum states?

These challenges have been appreciated for quite some time. Thouless once
claimed that it would be impossible to understand the transition from extended
states to localized states since there are uncharacterizable states lying
between them \cite[pg. 96]{Thou74a}. We call this \emph{Thouless' challenge}.
If we can build ``structured disorder'' materials as intermediates between
total randomness and exact crystalline periodicity, though, then we can begin
to study phenomena between extended and localized states and understand their
proper characterization. Here, we address Thouless' challenge by analyzing a
class of materials that span these structural extremes---the \emph{chaotic
crystals} discovered via X-ray diffraction studies \cite{Varn14a}---and
identifying a number of novel and counterintuitive material properties.

On the one hand, transport and localization of waves in disordered media has
been heavily studied for 60 years, starting with Anderson's pioneering results
\cite{Ande58a}. He pointed out that disorder can stop wave propagation and
diffusion depending on the degree of disorder. He analyzed the tight binding
model for a one-dimensional chain of random atoms, in which electrons hop
between neighboring sites. There, if the site energies are random, all
eigenstates localized \cite{Mot61a}.

On the other hand, in a periodic chain of atoms, Bloch's theorem says that
electrons can transit the lattice without backward scattering. Hence, in
exactly periodic lattices all wave states are extended states. Moreover,
conduction band structures emerge that are determined by lattice periodicity.

The last two decades witnessed several attempts to probe 
intermediate states via \emph{correlated disorder}, in which
random site energies are pairwise correlated. The first consequence was that
correlation renders localization very sensitive to wave energy. The second was
that long-range power-law correlation suppresses localization; this was
predicted theoretically \cite{Mour98a} and numerically verified \cite{Car02a}.
Moreover, with weak disorder, the relation between the Fourier transform of
disorder correlation and the localization versus energy dispersion was
established \cite{Izra99a}. By this, both localization suppression
\cite{Kuhl00a} and enhancement \cite{Kuhl08a} can be realized by designing
correlation with certain desired, if unphysical, features.

Correlated disorder definitely falls in between complete randomness and pure
periodicity. However, as implemented, it does not allow one to systematically
sweep between the structural extremes. The following puts forward the approach
of \emph{structured disorder} to build materials that allow this---their
lattices being an amalgam of periodic structure and disorder generated by \eMs,
a class of hidden Markov model \cite{Crut88a,Crut12a}. Lattices are formed from
their realizations which determine site energies across the lattice randomly,
but also in a way that gives a systematic dependence on the lattice ``history'' of
preceding energies. This dependence directly controls the degree and character
of material correlation. Tuning \eM\ parameters, then, one can sweep from a
periodic lattice to maximum-disorder structure and on to another periodic
lattice. The chaotic crystals so formed are generic in the sense that all
possible material lattices can be generated; specifically, stochastic lattices
of all different admixtures of periodicity and randomness can be exhaustively
enumerated \cite{John10a}. Constructively, the \eMs\ give a direct
specification, one that is local and so implementable, of how to assemble
chaotic crystals.

The following first introduces structured-disorder materials by showing how to
generate 1D atomic lattices from an \eM. It then studies the wavefunction
states that arise when sweeping through states intermediate between pure
periodicity and maximum disorder. Their nature is analyzed via the energy
spectrum of localization. We find materials in which highly enhanced localized
states and highly delocalized states coexist, even within very narrow energy
ranges. Moreover, at certain energies, a counterintuitive relation emerges
between degrees of disorder and localization. In addition, the density of
states for these chaotic crystals becomes richly structured, when sweeping from
complete randomness to correlated disorder. Previous studies had shown that,
with disorder, sharp band-boundaries do not persist and they bleed into band
gaps of the unperturbed system. In contrast, we find sharp band-boundaries
persisting in the presence of disorder and an abrupt emergence of positive
density of states in band gaps, rather than the density of states smoothly
stretching across band edges.

\paragraph*{Tight Binding Model}
Following Anderson, we consider the tight-binding model of a one-dimensional
material \cite{Ande58a}: an electron at energy $E$ moves through a chain of
atomic pseudopotentials (energies $\epsilon_n$) with one orbital per atomic
site $\ket{n}$. Physical properties of such a lattice are given by the
wavefunction $\Psi = \sum_n \psi_n \ket{n}$ determined by Schr\"odinger's
equation, which we solve iteratively:
\begin{align}
\psi_{n-1} + \psi_{n+1} = (E - \epsilon_n) \psi_n
  ~.
\label{eq:SchrodingerEqn}
\end{align}
(See Supplementary Materials for a review.)

If the chain is a perfectly regular crystal, all site energies $\epsilon_n$ are
uniform, forming a trivially periodic pattern of pseudopotentials on the
lattice. The solution to \cref{eq:SchrodingerEqn} is the Bloch wave $\psi_n =
e^{ikn}u_n$, with the periodic function $u_n$ indicating the wavefunction is
extensive over the whole lattice with a phase prefactor $e^{ikn}$. If material
disorder exists, in contrast, energies $\epsilon_n$ are randomly distributed,
say according to a Gaussian. Then, all eigenstates of \cref{eq:SchrodingerEqn}
are spatially localized. In other words, the probabilities $|\psi_n|^2$ in some
region dominate over those at other sites outside the region. And so, the
wavefunction is trapped there and there can be no electron transport.

\begin{figure*}[!htbp]     
\includegraphics[width=2.0\columnwidth]{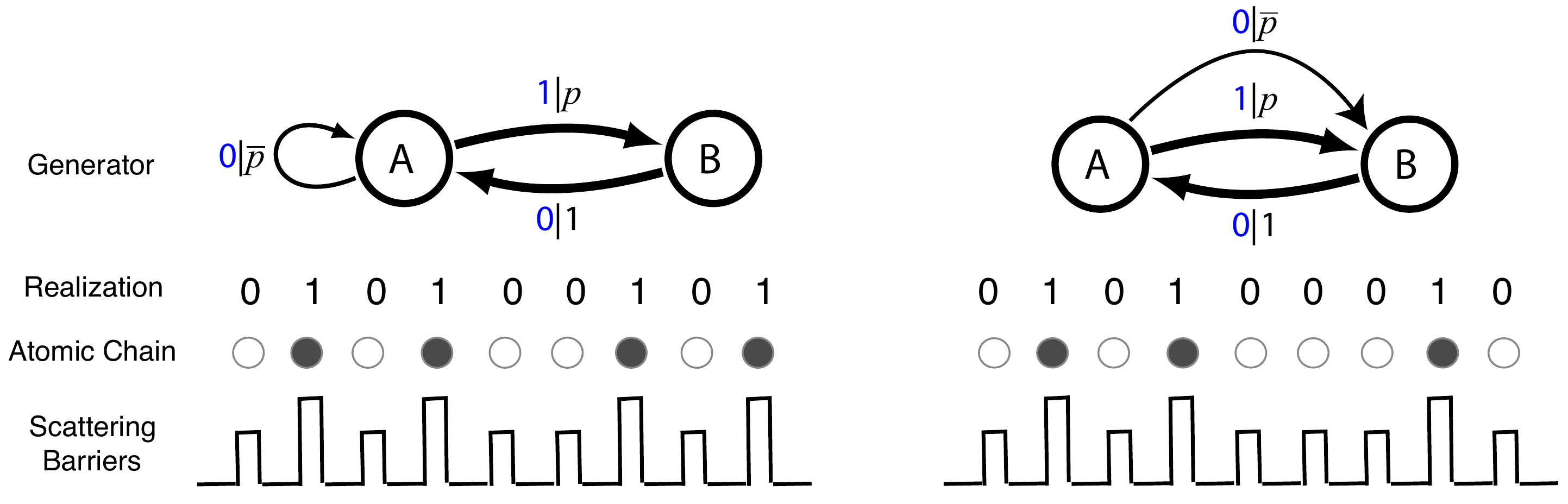}
\caption{Constructing structurally complex materials:
  (left) Golden Mean chaotic crystal---(top) Golden Mean Process \eM\ with two
  states $A$ and $B$ and three transitions labeled $x|p$ for taking the
  transition with probability $p$ (or, $\bar{p} = 1 - p$) and emitting symbol $x
  \in \{0, 1\}$, (bottom) Resulting lattice of scattering barriers (or
  pseudopotentials). (right) Noisy Period-$2$ chaotic crystal---(top) Noisy
  Period-$2$ Process \eM, (bottom) Resulting lattice of scattering barriers.
  White atom site energies $\epsilon_0 = 0.0$ and black atom $\epsilon_1 = 0.5$.
  }
\label{fig:ChaoticXtals}
\end{figure*}

Disorder also changes the material's energy spectrum---eigenvalues of
\cref{eq:SchrodingerEqn}. However, if the chain is a perfect crystal with $l$
atoms per cell, the energy spectrum forms a \emph{band structure} with $l$
conduction bands that support transport. Each band has sharp edges associated
with an abrupt change in the density of states (DoS). In the single-band
structure that appears for $l=1$, the allowed energy values are distributed in
an ordered pattern in the region $E \in (-2,2)$. For the familiar Gaussian
disordered chain, in contrast, there is no periodicity so that only a single
band emerges, with the energy amplitudes scattered randomly within. Moreover,
the sharp band edge with crystalline order becomes spread out by the disorder,
forming a rounded edge \cite{Lah08a}. The correlated disorder case also
produces a single band, no matter how the correlation statistics are
parametrized \cite{Car02a,Mour98a,Izra99a}. The band is squeezed, though,
forming very low DoS margins \cite{Car02a}. In short, in these materials
disorder smooths band edges.

\paragraph*{Structured Disorder}
Our challenge then is to systematically combine periodic lattices (especially
those with periodicity $l>1$) and lattices with different degrees of disorder.
We also want the resulting combination---which we call \emph{structured
disorder}---to be maximally expressive, so that every possible disordered
material can be described. To implement this, we regard the chain as a
stochastic process that evolves from left to right. Then we take the class of
\eMs\ as the stochastic process generator. \EMs\ form a class of hidden Markov
model (HMM) \cite{Rabi89a} whose statistical properties can be completely
analyzed in closed form \cite{Crut13a} and which can express every stochastic
process. In fact, as a representation \eMs\ allow for an exhaustive enumeration
of stochastic processes in terms of the latter's structure (memory) and
disorder (randomness) \cite{John10a}. Said simply, a given \eM\ specifies how
to construct a chaotic crystal and the enumeration provides a systematic way to
explore this large class of materials.

\Cref{fig:ChaoticXtals} shows how the construction works. The state transition
diagram at the top depicts the \eM\ that generates the \emph{Golden Mean
Process}---all binary strings are produced, except those with consecutive $1$s.
At each time step $t$, the generator is in one of two states $S_t \in \{A,
B\}$. A transition (labeled $x|p$) from current state to the next $S_{t+1}$ is
taken with the probability $p$ and the observable symbol $x_t \in \{0,1\}$ is
emitted. For example, if $S_0 = A$ and the \eM\ emits symbol $x_0 = 1$, then
system transitions to $S_1 = B$. At that point, a single transition emitting
$x_1 = 0$ and moving to state $S_2 = A$ is allowed. Thus, in the realizations
$x_0 x_1 x_2 \ldots x_{N-1}$ generated by this \eM\ consecutive $1$s are
forbidden.

We then construct a structured-disorder lattice from a given realization by
successively translating the observed symbols to atom types or, more literally,
to distinct local pseudopotentials. Let's map $x = 0$ to a ``white'' atom and
$x = 1$ to a ``black'' atom. Alternatively, if we were to work with the
Kronig-Penny model, we would map $0$, say, to a weaker scattering barrier and a
$1$ to a stronger scattering barrier. (This mapping is depicted in the bottom
two rows in \cref{fig:ChaoticXtals}.)

For the Golden Mean \eM\ with small $\overline{p}$, the cell $x_n x_{n+1} = 10$
dominates and only occasionally $0$s appear between them. We see that, in this
case, the lattice is a modification of a common dislocation type of
disorder. However, dislocated atoms are only of type $0$.

In addition to generating dislocation-disordered lattices, there are multiple
ways to combine periodic structures of $01$s with disorder $0$s. For example,
we can place disorder at every other site. The result is the chaotic crystal
generated by \emph{Noisy Period-$2$ Process}' \eM\ shown in
\cref{fig:ChaoticXtals}(b). In the cell $01$, only the first atom has a chance
to be a type-$0$ atom; the second is fixed to be a type-$1$ atom. Notably, this
type of structured disorder is realized in metamaterials with alternating
stacks of two layers \cite{Asat07a}. Though apparently closely related---at
first blush the \eMs\ seem nearly identical, only the $0|\bar{p}$ transition
has been redirected---the Golden Mean and Noisy Period-$2$ chaotic crystals
exhibit dramatically distinct localization phenomena, as we now show.

\paragraph*{Results}
To compare their material properties, we adapt transfer matrix products
\cite{Kram93a} to quantify the localization degree $\Lambda$, which is the
inverse Lyapunov exponent for the corresponding 2D map of wave vectors
$(\psi_n, \psi_{n-1}) \in \mathbb{R}^2$; see Supplementary Materials.  As a
main diagnostic, the following focuses on the \emph{localization energy
spectrum} $\Lambda(E)$: $\Lambda$ versus electron energy $E$.

\begin{figure}[!htbp]     
\includegraphics[width=1.0\linewidth]{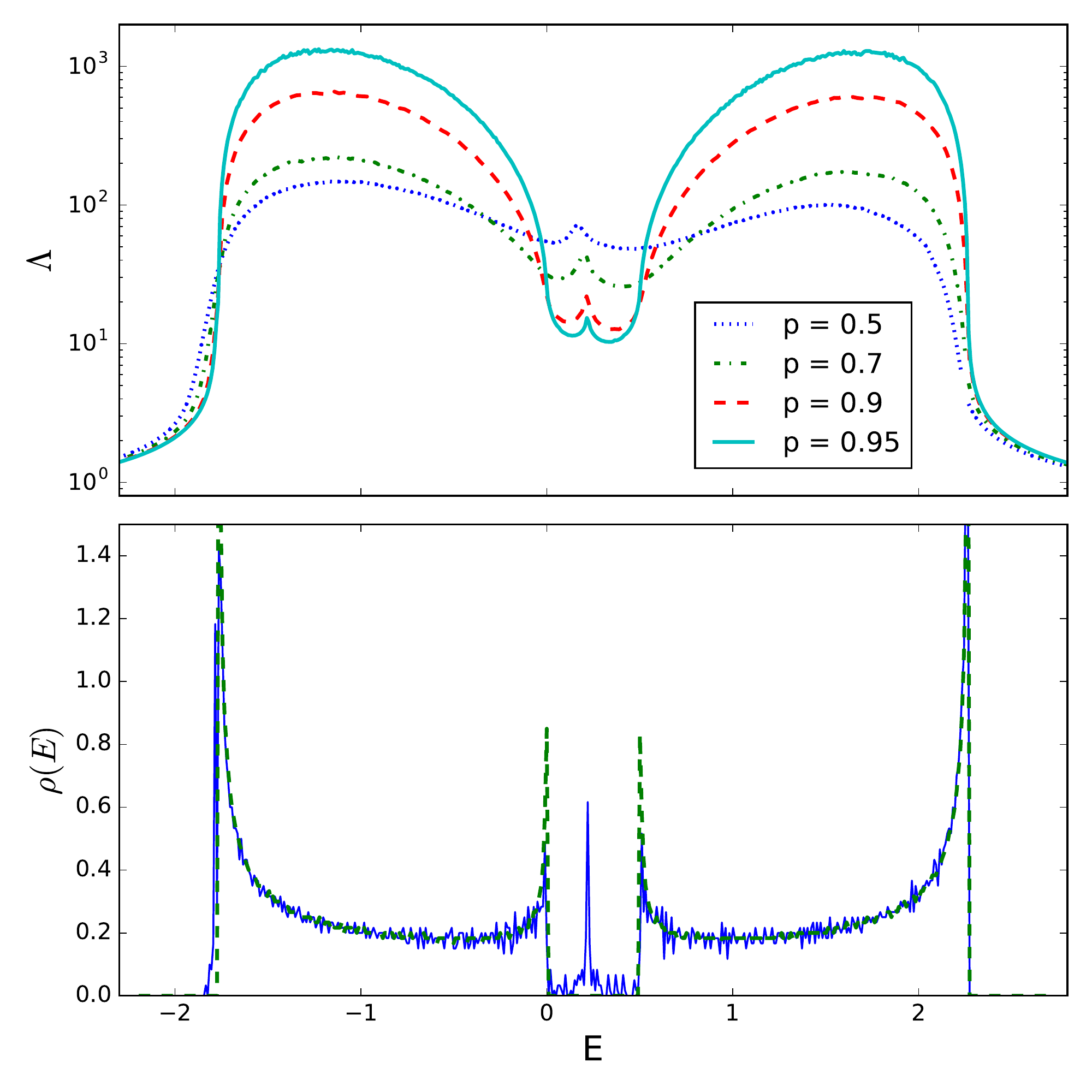}
\caption{Golden Mean chaotic crystal: (Top) Localization length $\Lambda$ versus
	energy $E$ for various disorder control settings $p$. $p=0.9$ indicates
	little degree of disorder, but the localization is very strong around the
	band gap at $E = 0$.
	(Bottom) Corresponding density of states $\rho(E)$. The green dashed line
	delineates the band boundaries for the unperturbed period-$2$ lattice.
	While the blue curve is for the Golden Mean chaotic crystal. Note the very
	sharp peak in $\rho(E)$ for new states in the gap. $\Lambda(E)$ is also
	positively correlated with this DoS feature.}
\label{fig:lambda_E_golden}
\end{figure}

For Golden Mean chaotic crystals, $p$ operates as a control for tuning from a
period-$2$ lattice to white-noise disorder. When $p \approx 1$, the period-$2$
pattern dominates the structured disorder. When $p \approx 0$, the
lattice is the simplest periodic structure of all type-$0$ atoms.
The lattice inherits no structure
from the period-$2$ tuning and the behavior appears as it does with very weak
white-noise disorder. (See
\cref{fig:lambda_E_golden}(Top).) For $p = 0.9$, even
though the lattice has very little disorder, the localization can be very
strong around energy range $E \in [0.0,0.5]$---the energy band gap
for the unperturbed period-$2$ lattice. Inside the two unperturbed bands,
localization is very weak. So, the Golden Mean chaotic crystal inherits the
band structure from the period-$2$ lattice. Decreasing $p$, this inheritance
weakens. And, for $p = 0.3$, the localization length versus energy curve is
basically flat as seen with Gaussian disorder.

\begin{figure}[!htbp]     
\includegraphics[width=1.0\linewidth]{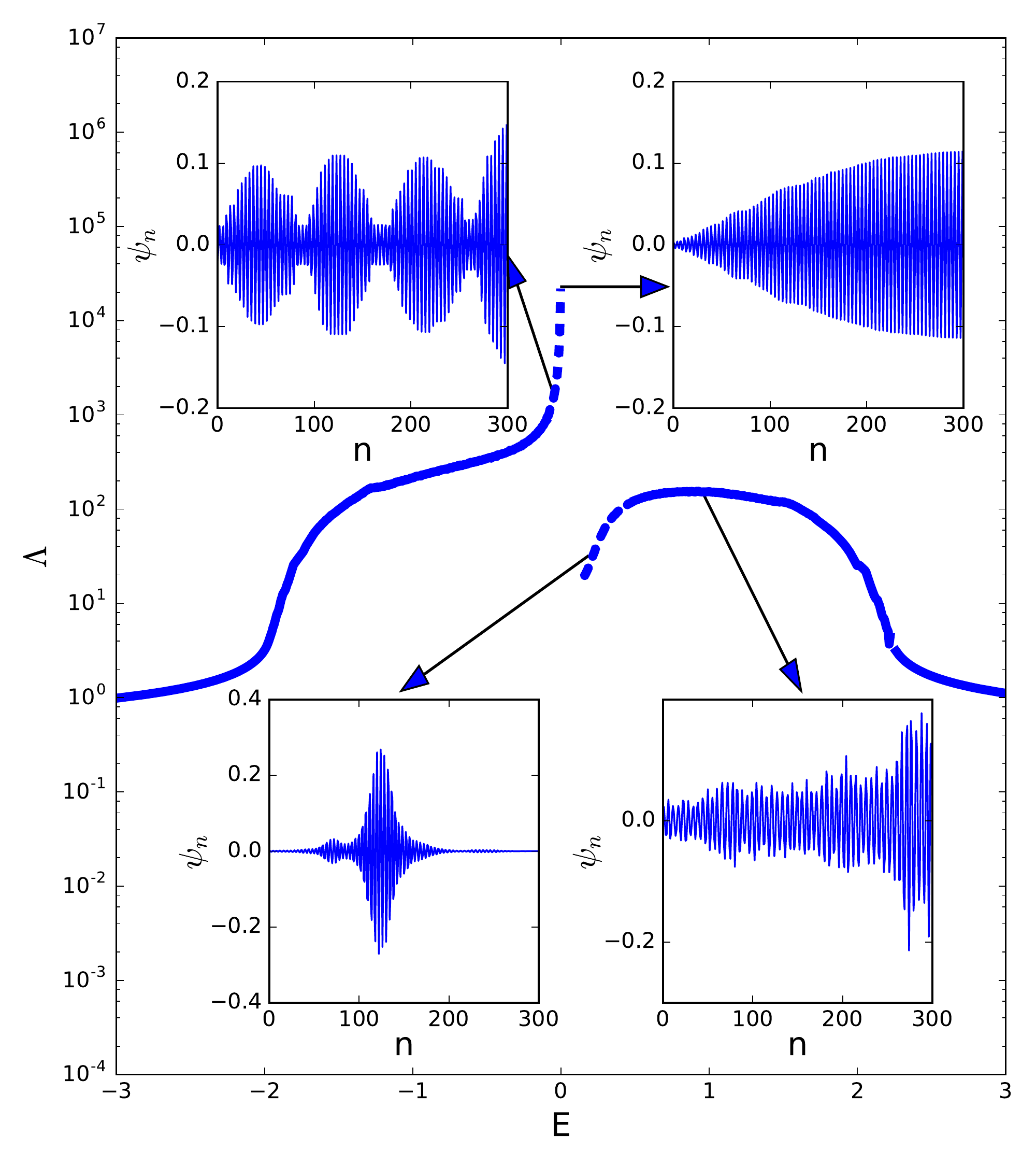}
\includegraphics[width=1.0\linewidth]{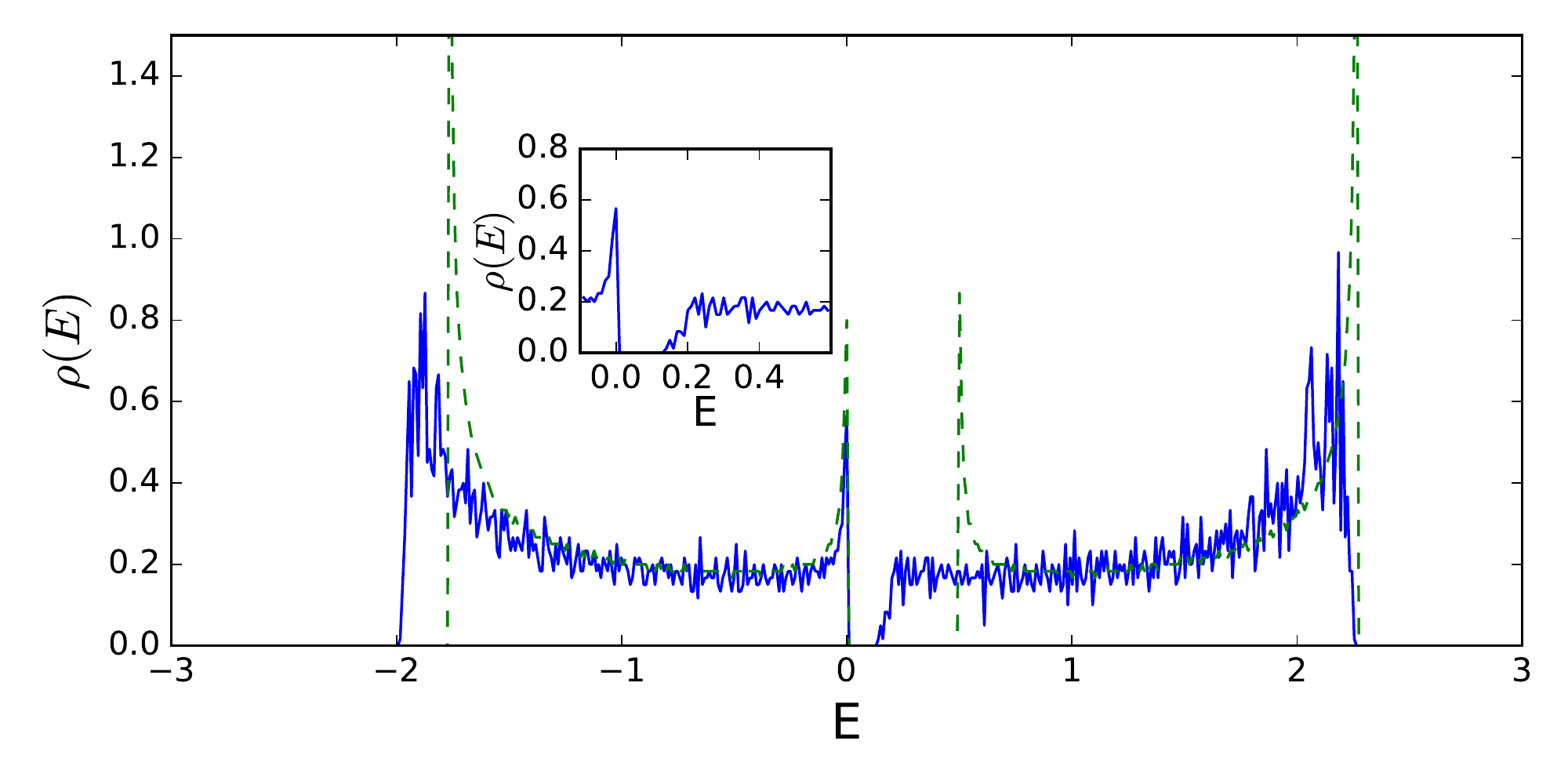}
\caption{Noisy Period-$2$ chaotic crystal with $p=0.5$: (Top) Localization
	length
	$\Lambda$ versus energy $E$ and typical wavefunctions at four different $E$.
	Left upper: Connected bump-like delocalized state at E=$-0.03$.
	Right upper: Linear expanding wave in a single bump at E=$-0.0004$.
	Left lower: Enhanced localized state at E=$0.23$.
	And, right lower: Typical localized state at E=$1.00$.
	(Bottom) Corresponding density of states $\rho(E)$. The green dashed line delimits
	the band boundaries that would be exhibited for the unperturbed period-$2$
	lattice. While the blue curve is for the structured disorder in the Noisy
	Period-$2$ chaotic crystal. Notably, the gap's left boundary persists
	without broadening. As a consequence, the DoS remains high in the
	unperturbed left band. We do observe delocalization at this energy, showing
	positive relation with the DoS. In the inset figure, a zoom-in display of 
  DoS near the band gap highlights this relation.
  }
\label{fig:lambda_E_noisy}
\end{figure}

Curiously, inherited band structure does not always disappear with increasing
disorder. For Noisy Period-$2$ chaotic crystals, periodicity $2$ persists for
any $p$ value, except $p=0$. In other words, only even or odd sites (depending
on the realization's beginning) can have black atoms, no matter how different
the realization is from the underlying period-$2$ lattice. With its different
underlying structure, the Noisy Period-$2$ crystal has a completely different
localization versus energy curve; see $\Lambda(E)$ in
\cref{fig:lambda_E_noisy}(Top). Localization is still strong inside the band gap.
However, at the right boundary of the left band, the localization length
increases and forms a singularity.

Delocalization here is due to the persistent periodicity in the Noisy
Period-$2$ Process. A typical delocalized wave is shown at the upper left
corner in \cref{fig:lambda_E_noisy}(Top). A significant feature is the strong
periodic modulation of the wave amplitude.  One sees that these ``bumps'' are
amplified and localized. With energy $E$ closer to the critical value $E_c =
0$, the bump width increases and localization weakens. At $E_c$, the wave
reduces to a single bump with linear expanding envelope; see the upper right
corner in \cref{fig:lambda_E_noisy}(Top). Hence, the Lyapunov coefficient $\gamma$
vanishes and localization is completely suppressed. Across the singularity
boundary, though, inside the unperturbed band gap, localization is greatly
enhanced. So, an abrupt transition from extensive state (conductor) to strongly
localized state (insulator) is achieved within an exceedingly narrow energy
range. Such abrupt transition features can be harnessed to design new 1D
layered structures for capturing and transporting electrons in desired
regions.

In addition to novel localization properties, nontrivial features arise in a
chaotic crystal's density of states around band boundaries and band gaps. With
normal disorder added to the period-$2$ lattice, the band structure only
smooths out at band boundaries. In the Golden Mean chaotic crystal, however,
new states emerge in the \emph{middle of the unperturbed gap}. Specifically, at
$p=0.95$ the DoS shows a sharp peak at energy $E=0.25$ with low density nearby;
see \cref{fig:lambda_E_golden}(Bottom). In other words, rather than shrinking
the band gap, an island of highly localized states emerges within the gap.
Corresponding to the DoS, the $\Lambda(E)$ also shows a peak at the same
energy. This is explained by Jones and Thouless's theorem on the relation
between localization level and DoS \cite{Her71a,Thou72a}. Specifically, in
exponentially localized states, the localization level at energy $E$ is
positively related to the number of states existing around $E$. For energy
inside the band gap, except the island, nearby states are fewer so that
localization is stronger. At the island energy, though, there are more nearby
states, so a peak in $\Lambda(E)$ is observed; though still highly localized
compared to the inside-band wavefunctions.

The emergence of localization islands suggests a novel way to manipulate a
material's conductivity. By tuning the Golden Mean's disorder, we can
turn-on a new ``band'' in the band gap that absorbs eigenstates previously at
conducting bands' edges. Thus, if the Fermi surface is at the gap's middle,
even small Golden Mean disorder will strongly suppress conductance. 

The Noisy Period-$2$ chaotic crystal, see \cref{fig:lambda_E_noisy}(Bottom),
exhibits the same positive relation between localization length $\Lambda(E)$
and the DoS. This time, though, the smoothing of the right band is the same as
that seen with normal disorder. However, the left band's boundary persists
exactly as if it was an unperturbed band boundary. The high DoS inside the
boundary also remains. Correspondingly, we observe delocalization around
this boundary. Notably, the Jones and Thouless's theorem does not apply here
since it is valid only for exponentially localized states and not for the new
``modulated-bumps'' wavefunction.

\paragraph*{Conclusion}
Our strategy in the preceding is simple to state: We revisited Anderson
localization theory and band-gap theory from the perspective of computational
mechanics. The latter gave a systematic and quantitative view of the spectrum
of structural complexity as one looks across processes with different mixtures
of structure and disorder. Its lesson is that the most structurally complex
processes occur in the intermediate regime of disorder---not exactly periodic,
not utterly random. The result of this synthesis was several fold. First, we
introduced a new family of chaotic crystal materials and gave a constructive
specification (the \eM) for how to assemble them. Second, our explorations
found that this family exhibits a range of novel, perhaps surprising
properties. Strong localization emerges from weak disorder. Enhanced
localization and transport coexist within narrow energy ranges. Islands of
localized states arise within band gaps. And, sharp band-gap boundaries persist
in the presence of substantial disorder. Practical consequences immediately
suggest themselves.

Localization enhancement is a prospect for insulator material design. For a
period-$2$ lattice with a single valence electron, the Fermi-surface is in the
band gap, resulting in an insulator. Adding Golden Mean disorder leads to a DoS
island just around the Fermi-surface. Since states in the island are strongly
localized, a superinsulator is constructed with very little disorder.  Also,
for Noisy Period-$2$ disorder, the gap between extensive states and strongly
localized states can be very small. There, an external electric field will
cause electronics to drift, via the transition through the narrow band gap. Once
transmitted, though, the electrons are localized again. The induced electron
drift and sudden-stop phenomena provides yet another kind of material property
control.

How might one test for these predicted phenomena? Direct experimental probes of
Anderson localization in electronic waves is notoriously challenging, owing to
electron-electron interactions. However, nanowires provide a possible 1D
setting for observing electron localization \cite{garc95a,gome05a}. There are
also other arenas in which the wave phenomena predicted here should arise. For
example, disordered photonic materials and waveguides have led to successful
experimental observation of localization of light and microwaves. All of our
predictions can be verified on such systems with designed materials and
waveguide geometries, requiring relatively simple experimental setups.

The authors thank Santa Fe Institute for its hospitality during visits and
thank Phil Anderson and Paul Riechers for helpful conversations. JPC is an SFI
External Faculty member. This material is based upon work supported by, or in
part by, the John Templeton Foundation grant 52095, the Foundational Questions
Institute grant FQXi-RFP-1609, and the U. S. Army Research Laboratory and the
U. S. Army Research Office under contracts W911NF-13-1-0390 and
W911NF-13-1-0340.

\bibliography{chaos}

\begin{thebibliography}{10}

\bibitem{Thou74a}
D.~J. Thouless.
\newblock Electrons in disordered systems and the theory of localization.
\newblock {\em Phys. Reports}, 13(3):93--142, 1974.

\bibitem{Varn14a}
D.~P. Varn and J.~P. Crutchfield.
\newblock Chaotic crystallography: {How} the physics of information reveals
  structural order in materials.
\newblock {\em Curr. Opin. Chem. Eng.}, 7:47--56, 2015.

\bibitem{Ande58a}
P.~W. Anderson.
\newblock Absence of diffusion in certain random lattices.
\newblock {\em Phys. Rev.}, 109:1492--1505, Mar 1958.

\bibitem{Mot61a}
N.~F. Mott and W.~D. Twose.
\newblock The theory of impurity conduction.
\newblock {\em Adv. Physics}, 10(38):107--163, 1961.

\bibitem{Mour98a}
F.~A. B.~F. de~Moura and M.~L. Lyra.
\newblock Delocalization in the {1D Anderson} model with long-range correlated
  disorder.
\newblock {\em Phys. Rev. Lett.}, 81:3735--3738, Oct 1998.

\bibitem{Car02a}
P.~Carpena, P.~Bernaola-Galvan, P.~Ch. Ivanov, and H.~E. Stanley.
\newblock Metal-insulator transition in chains with correlated disorder.
\newblock {\em Nature}, 418(6901):955--959, 08 2002.

\bibitem{Izra99a}
F.~M. Izrailev and A.~A. Krokhin.
\newblock Localization and the mobility edge in one-dimensional potentials with
  correlated disorder.
\newblock {\em Phys. Rev. Lett.}, 82:4062--4065, May 1999.

\bibitem{Kuhl00a}
U.~Kuhl, F.~M. Izrailev, A.~A. Krokhin, and H.-J. Stockmann.
\newblock Experimental observation of the mobility edge in a waveguide with
  correlated disorder.
\newblock {\em App. Phys. Lett.}, 77(5):633--635, 2000.

\bibitem{Kuhl08a}
U.~Kuhl, F.~M. Izrailev, and A.~A. Krokhin.
\newblock Enhancement of localization in one-dimensional random potentials with
  long-range correlations.
\newblock {\em Phys. Rev. Lett.}, 100:126402, Mar 2008.

\bibitem{Crut88a}
J.~P. Crutchfield and K.~Young.
\newblock Inferring statistical complexity.
\newblock {\em Phys. Rev. Let.}, 63:105--108, 1989.

\bibitem{Crut12a}
J.~P. Crutchfield.
\newblock Between order and chaos.
\newblock {\em Nature Physics}, 8(January):17--24, 2012.

\bibitem{John10a}
B.~D. Johnson, J.~P. Crutchfield, C.~J. Ellison, and C.~S. McTague.
\newblock Enumerating finitary processes.
\newblock arxiv.org:1011.0036.

\bibitem{Lah08a}
Y.~Lahini, A.~Avidan, F.~Pozzi, M.~Sorel, R.~Morandotti, D.~N. Christodoulides,
  and Y.~Silberberg.
\newblock Anderson localization and nonlinearity in one-dimensional disordered
  photonic lattices.
\newblock {\em Phys. Rev. Lett.}, 100(1):013906, 2008.

\bibitem{Rabi89a}
L.~R. Rabiner.
\newblock A tutorial on hidden {Markov} models and selected applications.
\newblock {\em IEEE Proc.}, 77:257, 1989.

\bibitem{Crut13a}
J.~P. Crutchfield, P.~Riechers, and C.~J. Ellison.
\newblock Exact complexity: {Spectral} decomposition of intrinsic computation.
\newblock {\em Phys. Lett. A}, 380(9-10):998--1002, 2016.

\bibitem{Asat07a}
A.~A. Asatryan, L.~C. Botten, M.~A. Byrne, V.~D. Freilikher, S.~A. Gredeskul,
  I.~V. Shadrivov, R.~C. McPhedran, and Y.~S. Kivshar.
\newblock Suppression of {Anderson} localization in disordered metamaterials.
\newblock {\em Phys. Rev. Lett.}, 99(19):193902, 2007.

\bibitem{Kram93a}
B.~Kramer and. A.~MacKinnon.
\newblock Localization: Theory and experiment.
\newblock {\em Rep. Prog. Physics}, 56(12):1469, 1993.

\bibitem{Her71a}
D.~C. Herbert and R.~Jones.
\newblock Localized states in disordered systems.
\newblock {\em J. Phys. C}, 4(10):1145, 1971.

\bibitem{Thou72a}
D.~J. Thouless.
\newblock A relation between the density of states and range of localization
  for one dimensional random systems.
\newblock {\em J. Phys. C}, 5(1):77, 1972.

\bibitem{garc95a}
N.~Garcia, U.~Landman, W.~D. Luedtke, E.~N. Bogachek, and H.P. Cheng.
\newblock Properties of metallic nanowires: from conductance quantization to
  localization.
\newblock {\em Science}, 267:1793, 1995.

\bibitem{gome05a}
C.~G{\'o}mez-Navarro, P.~J. De~Pablo, J.~G{\'o}mez-Herrero, B.~Biel, F.~J.
  Garcia-Vidal, A.~Rubio, and F.~Flores.
\newblock Tuning the conductance of single-walled carbon nanotubes by ion
  irradiation in the {Anderson} localization regime.
\newblock {\em Nat. Materials}, 4(7):534--539, 2005.

\end{thebibliography}

\clearpage
\begin{center}
\large{Supplementary Materials}\\
\emph{Islands in the Gap:\\
Intertwined Transport and Localization in\\
Structurally Complex Materials\\
X. Lei, D. P. Varn, and J. P.  Crutchfield}
\end{center}

%%%%%%%%%% Supplemental Materials %%%%%%%%%%
%%%%%%%%%% Prefix a "S" to all equations, figures, tables and reset the
%counter
%%%%%%%%%%%
\setcounter{equation}{0}
\setcounter{page}{1}
\makeatletter
\renewcommand{\theequation}{S\arabic{equation}}

\section{Tight Binding Model}
Consider the tight-binding model of a one-dimensional material \cite{Ande58a}.
Its Hamiltonian describes an electron at energy $E$ moving through a chain of
atomic pseudopotentials with one orbital per atomic site:
\begin{align}
\widehat{H} = \sum_n^{N-1} \epsilon_n \ket{n} \bra{n} 
  + \sum_{\langle n,m \rangle} t \ket{n} \bra{m}
  ~,
\label{eqa:TBM}
\end{align}
where $\ket{n}$ is the orbital at site $n$ with energy $\epsilon_n$ and the
second term corresponds to the hopping between nearest neighbor sites $\langle
n,m \rangle$ with hopping rate $t$. For simplicity, we investigate
\emph{diagonal disorder} focusing on effects arising from disorder in site
energies $\epsilon_n$, while regarding hopping rate $t$ as uniform between
neighbors. We set the rate $t=1$ and define it as the unit for site energy
disorder.

Physical properties of such a lattice are determined by a Schr\"odinger wave
equation specified by the Hamiltonian in \cref{eqa:TBM}. For the tight-binding
model, we expand the electron's eigenfunction $\Psi$ in terms of $\ket{n}$:
\begin{align}
\Psi = \sum_n^{N-1} \psi_n \ket{n}
  ~.
\label{eq:Psi}
\end{align}
We solve for the wavefunction via an iterative relation between neighboring
sites:
\begin{align*}
\psi_{n-1} + \psi_{n+1} = (E - \epsilon_n) \psi_n
  ~,
\end{align*}
where we must specify the initial condition $\psi_0$ and $\psi_1$ at, say, the
lattice's left end. The lattice's right end is site $N-1$.

\section{Numerical Simulation and Quantifiers}

To compare their material properties, we adapt transfer matrix products
\cite{Kram93a} to quantify the localization degree. Rewriting the tight-binding
model of \cref{eq:SchrodingerEqn} into a map $T_n$ of vectors $(\psi_n,
\psi_{n-1}) \in \mathbb{R}^2$ gives:
\begin{align}
\left[ \begin{array}{c} \psi_{n+1} \\ \psi_n \end{array} \right] = \begin{bmatrix} E - \epsilon_n & -1 \\ 1 & 0 \end{bmatrix} \times \left[ \begin{array}{c} \psi_n \\ \psi_{n-1} \end{array} \right]
  ~.
\end{align}
The product $Q_N = \prod_{n=1}^{N} T_n$ controls whether the wave amplifies or
attenuates going from left to right, from $n = 0$ to $n = N-1$. The more rapid
the divergence, for example, the stronger the localization, assuming the
localized packet's peak occurs at the far right.

Interpreting the wavefunction transfer matrix as a dynamical system, we
employ the Lyapunov coefficient $\gamma$ to measure the exponential rate of
divergence per site:
\begin{align}
\gamma = \lim_{N \to \infty} \log \left( \frac{\|Q_N\|_2}{N} \right)
  ~.
\end{align}
$\Lambda = \gamma^{-1}$ is a wavefunction's localization characteristic length.
Small $\Lambda$ indicates strong localization and no transport. As a main
diagnostic, the following focuses on the \emph{localization energy spectrum}
$\Lambda(E)$: $\Lambda$ versus electron energy $E$.

Following in the long tradition of numerical simulation of disorder, we study
$\Lambda(E)$ on a chain of $10^6$ sites. This gives excellent convergence
across different structured disorder realizations. For white atoms, we set the
site energy to $\epsilon_0 = 0.0$, while $\epsilon_1 = 0.5$ for black atoms. In
this case, the period-$2$ lattice has an energy gap of width $0.5$. This allows
us to easily observe phenomena arising inside the gap. For the density of
states, we calculate the eigenvalue distribution of the Hamiltonian matrix for
a system with $6 \times 10^3$ sites; which also achieves excellent
convergence.

\end{document}